\documentstyle[12pt,aps,prd,preprint]{revtex}
\begin{document}
\title{Cosmic Censorship: The Role of Quantum Gravity}
\author{Shahar Hod and Tsvi Piran}
\address{The Racah Institute for Physics, The
Hebrew University, Jerusalem 91904, Israel}
\date{\today}
\maketitle

\begin{abstract}
  The cosmic censorship hypothesis introduced by Penrose thirty years
  ago is still one of the most important open questions  
  in {\it classical} general relativity. In this essay we put forward the
  idea that cosmic censorship is intrinsically a {\it quantum gravity}
  phenomena.  To that end we construct a gedanken experiment in which
  cosmic censorship is violated within the purely {\it classical}
  framework of general relativity. We prove, however, 
that {\it quantum} 
effects restore the validity of the conjecture. This suggests that 
classical general relativity is inconsistent and that cosmic
censorship might be enforced only by a quantum theory of gravity.
\end{abstract}
\bigskip

Spacetime singularities that arise in gravitational collapse are
always hidden inside of black holes. 
This is the essence of the (weak) cosmic censorship conjecture, 
put forward by Penrose thirty years ago \cite{Pen}. 
The conjecture, which is widely believed to be true, has become one of the corner stones of 
general relativity. Moreover, it is being envisaged as a
basic principle of nature.  However, despite the flurry of activity
over the years, the validity of this conjecture is still an open
question (see e.g., \cite{Wald1,His} for reviews). 

The destruction of a black hole's  event horizon is ruled out by this
principle because it would expose the inner singularities to distant
observers. Moreover, the horizon area of a black hole, $A$, 
is associated with an entropy $S_{BH}=A/4\hbar$ (we use $G=c=1$). 
Therefore, without any obvious physical mechanism to compensate for the
loss of the black-hole enormous entropy, 
the destruction of the black-hole event horizon would 
violate the generalized second law (GSL) of thermodynamics
\cite{Beken1}. For these two reasons, any process which seems, at
first sight, to remove 
the black-hole 
horizon is expected to be unphysical. For the advocates of the cosmic
censorship principle the task remains to find out how such candidate
processes eventually fail to remove the horizon.

The main goal of this essay is to put forward the idea that the
stability of the black-hole horizon, and  
the {\it cosmic censorship} 
principle are intrinsically {\it quantum} phenomenon. To
that end, we construct a gedanken experiment in which cosmic
censorship is being violated within the purely {\it classical}
framework of general relativity. We prove, however, that {\it quantum}
effects save the cosmic censorship principle.

One of the earliest attempts to eliminate the horizon of a black hole
is due to Wald \cite{Wald2}. As is well-known, the
Reissner-Nordstr\"om metric with $M < Q$ (where $M$ and $Q$ are the
mass and charge) does not contain an event
horizon, and it therefore describes a naked singularity. Wald tried
to ``over-charge'' an extremal black hole (characterized by $Q=M$)
by dropping into it a charged test particle whose charge-to-mass 
ratio is larger
than unity.  Wald considered the specific case of a particle
which starts falling from spatial {\it infinity} (thus, the particle's
energy-at-infinity is larger than its rest mass).  He has shown that
this attempt to ``over-charge'' the black hole would fail 
because of the Coulomb potential barrier surrounding the black hole.

A more `dangerous' version of Wald's 
original gedanken
experiment is one in which the charged particle is {\it slowly}
lowered towards the black hole.  In this case, the energy delivered to
the black hole (the part contributed by the body's rest mass, see
below) can be {\it red-shifted} by letting the assimilation point
approach the black-hole horizon. On the other hand, the particle's
charge is not redshifted by the gravitational field of the black hole.
At a first sight 
 the particle [with arbitrarily small (redshifted) mass-energy] is not hindered 
from entering the black hole and
removing its horizon, thereby violating cosmic censorship.


Consider a charged body of rest mass $\mu$, charge $q$, and proper
radius $b$, which is {\it slowly} descent into a (near extremal) black
hole.  The total energy $\cal E$ (energy-at-infinity) of the body in a
black-hole spacetime is made up of three contributions: $ 1) \ {\cal
  E}_0=\mu (g_{00})^{1/2}$, the energy associated with the body's mass
(red-shifted by the gravitational field); $ 2) \ {\cal
  E}_{elec}=eQ/r$, the electrostatic interaction of the charged body
with the external electric field; and $ 3) \ {\cal E}_{self}$, the
gravitationally induced self-energy of the charged body.


The physical origin of the third contribution, ${\cal E}_{self}$, is the
distortion of the charge's long-range Coulomb field by the spacetime
{\it curvature}. This results in 
a repulsive (i.e., directed away
from the black hole) self-force in the black-hole background. A variety of techniques have been
used to demonstrate this effect in black-hole spacetimes. In
particular, the contribution of this effect to the particle's (self)
energy in the Reissner-Nordstr\"om background 
is ${\cal E}_{self}=Mq^2/2r^2$ \cite{ZelFro}.

The total energy of a 
charged particle at 
a proper
distance $\ell$ ($\ell \ll r_+$) above the horizon is  given by:
\begin{equation}\label{Eq1}
{\cal E}(\ell)={{\mu \ell (r_+-r_-)} \over {2{r^2_+}}} +{{qQ} \over {r_{+}}}
-{{qQ\ell^2(r_{+}-r_{-})} \over {4{r^4_+}}}+ {{Mq^2} \over {2{r^2_+}}}\  ,
\end{equation}
where $r_{\pm}$ are the
locations of the black-hole (event and inner) horizons. This expression is actually the effective
potential governing the motion of a charged body in the black-hole
spacetime. Provided $qQ>0$, it has a {\it maximum} located at
$\ell=\ell^*(\mu,q;M,Q)=\mu r^2_+ /qQ$.  


The most challenging situation for the cosmic censorship conjecture
occurs when the charge-to-energy ratio of
the captured particle is as large as possible. This can be achieved if 
one {\it slowly} lowers the body as close to the horizon as
possible. However, an object suspended in the vicinity
of a black hole is actually accelerated by virtue of its being
prevented from falling freely along a geodesic. As first pointed
out by Unruh and Wald \cite{UnWa}, the object would  
feel isotropic thermal radiation, the well-known Unruh radiance
\cite{UnWa}. As a consequence, buoyancy in the
radiative black-hole environs will {\it prevent} lowering the object
{\it slowly} all the way 
down to the horizon. 
It will float at a  proper height $\ell=b$, 
almost {\it touching} the  horizon. 
The energy
(energy-at-infinity) delivered to the black hole 
is minimized when the object is released to fall in from this
flotation 
point \cite{UnWa}. One should therefore evaluate $\cal E$ at the point $\ell=b$. 

An assimilation of the charged object results with a change $\Delta
M={\cal E}$ in the black-hole mass and a change $\Delta Q=q$ in its
charge. The condition for the black hole to preserve its integrity
after the assimilation of the body is: 
\begin{equation}\label{Eq2}
q+Q \leq M+{\cal E}\  .
\end{equation}
Substituting ${\cal E}={\cal E}_0+{\cal E}_{elec}+{\cal E}_{self}$
from Eq. (\ref{Eq1}) one finds a necessary and sufficient condition
for removal of the black-hole horizon:

\begin{equation}\label{Eq3}
(q-\varepsilon)^2+{{2\varepsilon} \over M} \Bigg (\mu b -q^2 -{{qb^2} \over
  {2M}} \Bigg) +{{q\varepsilon^2} \over M} < 0\  ,
\end{equation}
where $r_{\pm} \equiv M \pm \varepsilon$. The expression on the l.h.s. of
Eq. (\ref{Eq3}) is minimized
for $q = \varepsilon +O(\varepsilon^2/M)$, yielding

\begin{equation}\label{Eq4}
2\mu b -q^2 -qb^2/M <0\  ,
\end{equation}
as a necessary and sufficient condition for elimination of the
black-hole horizon. This condition (together with the requirement 
$b \leq \ell^*$, the case $\ell^* < b$ is discussed below) simply
implies 
that the charged object
must be smaller than its classical radius. However, any charged body
which respects the weak (positive) energy condition (i.e., it does not
have a region of negative energy density in it) must be larger than
its classical radius. We therefore conclude that the black-hole
horizon cannot be removed by an assimilation of such a charged body --
cosmic censorship is upheld !  

We emphasize that the {\it quantum} buoyancy due to the
Unruh-Wald radiance is a crucial ingredient in this analysis. Without
it one could have {\it slowly} lowered the object down to the horizon (thereby
completely redshifting its mass-energy), and it would have been possible to
violate cosmic censorship (together with a violation of the
GSL). 


If the radius of the charged object is larger than $\ell^*$, then it
must have a minimal energy of ${\cal E}_{min}={\cal E}(\ell^*)$ 
in order to overcome the potential barrier, and to be
captured by the black hole (recall that the effective potential
barrier has a maximum located at $\ell=\ell^*$). 
This is also true for any charged object
which is released to fall freely from $\ell > \ell^*$, in which case the
Unruh-Wald buoyancy can be made arbitrarily negligible (if $\ell >> b$). 
Taking cognizance of
Eq. (\ref{Eq4}) (with $b$ replaced by $\ell^*$) we find that a necessary and sufficient condition for
removal of the black-hole horizon in this case is $2\mu \ell^*
-q^2-q\ell^{*2}/M<0$, or equivalently,

\begin{equation}\label{Eq5}
{\mu^2}/{q^3} < E\  ,
\end{equation}
where $E = Q/{r^2_+}=M^{-1}+O(\varepsilon /M^2)$ is the
black-hole electric field in the vicinity of its horizon.


The assimilation of a charged object by a
charged black hole satisfying condition  (\ref{Eq5}) would
{\it violate} the cosmic censorship conjecture. 
There is no classical effect that could prevent this. 
However, Schwinger discharge (vacuum polarization), a 
purely {\it quantum} effect sets an 
upper bound to the black-hole electric field and  
saves
cosmic censorship. Pair-production 
of the {\it lightest} charged particles imply a maximal
(critical) electric field: 
$E \leq E_c \equiv {{\pi {m^2_e}}/{|e| \hbar}}$, where $m_e$ and $e$
are the rest mass and charge of the
electron, respectively. 
A necessary 
condition for a violation of the cosmic censorship conjecture within
the framework of a quantum theory is the existence of
a charged object which satisfies the inequality
\begin{equation}\label{Eq6}
q^3E_c/{\mu^2} > 1\  .
\end{equation}
Obviously, the most dangerous threat to the integrity of the black
hole is imposed by the electron, which has the largest charge-to-mass
ratio in nature. However, even the electron itself satisfies the
relation $q^3E_c/{\mu^2}=\pi \alpha < 1$ (where $\alpha=e^2/\hbar
\simeq 1/137$ is the fine structure constant), and thus it {\it
  cannot} remove the black-hole horizon.  Atomic nuclei, the densest
composite charged objects in nature satisfy the relation
$q^3E_c/{\mu^2} \lesssim 10^{-7}$ and are therefore absolutely
harmless to the black hole. Thus, vacuum polarization (Schwinger
discharge of the black hole) insures 
the integrity of the black hole. Without this {\it quantum} mechanism one
could have removed the black-hole horizon, thereby exposing a naked
singularity. It seems that nature has ``conspired'' to prevent this.

We have shown that two purely quantum effects -- Unruh radiation and
Schwinger discharge are essential for saving cosmic censorship. Is there any
classical effect that we have neglected that could 
 save cosmic
censorship ? In the analysis presented so far we have assumed that
corrections to the metric do not effect the particle's energy to order
$O(q^2)$. A correction of this order would modify condition (3) in
such a way that it will be either always satisfied (in which case it
would be always possible to violate cosmic censorship regardless of
quantum effects) or that it will always be violated (making cosmic
censorship viable on a classical level).  However, we expect that
there is no correction to the particle's energy of order $O(q^2)$
(except of the self-energy term ${\cal E}_{self}$, which we have already
taken into account). In
our analysis we have considered the motion of the particle on the
unperturbed metric.  In the other extreme case the particle would move
on the modified 
 metric with the corrected parameters $M \rightarrow
M+ q Q/r_+ + O(q^2/M)$ and $Q \rightarrow Q +q$.  In this final metric
corrections of order $O(q)$ are canceled out, and the metric 
is
corrected only to the order of $O(q^2)$, thereby yielding only 
a correction of order
$O(q^3)$ to particle's energy.



Although the question of whether cosmic censorship holds remains very
far from being settled, we find from this gedanken experiment that the
black-hole event horizon may be {\it classically} unstable while
absorbing charged objects. This suggests that the purely classical
laws of general relativity do not enforce cosmic censorship.  However,
{\it quantum} effects insure the stability of the black-hole event
horizon, and thereby restore the validity of the cosmic censorship
principle. We thus conclude that the cosmic censor must be cognizant
of {\it quantum gravity}.
 
\bigskip
\noindent
{\bf ACKNOWLEDGMENTS}
\bigskip

It is a pleasure to thank Jacob D. Bekenstein for stimulating discussions. 
This research was supported by a grant from the Israel Science Foundation.

\end{document}